\documentstyle[12pt]{article}
\begin{document}
\begin{flushright}
GUTPA/98/12/03\\
\end{flushright}
\vskip .1in

\begin{center}

{\Large \bf Where does Flavour Mixing come from?}

\vspace{50pt}

{\bf J.L. Chkareuli}

\vspace{6pt}

{ \em  International Centre for Theoretical Physics,
34100 Trieste, Italy  and\\
Institute of Physics, Georgian Academy of Sciences, 380077 Tbilisi, Georgia\\}

\vspace{18pt}

{\bf C.D. Froggatt}

\vspace{6pt}

{ \em Department of Physics and Astronomy\\
 Glasgow University, Glasgow G12 8QQ,
Scotland\\}
\end{center}

\section*{ }
\begin{center}
{\large\bf Abstract}
\end{center}
We argue that flavour mixing, both in the quark and charged lepton
sector, is basically determined by the lightest family mass
generation mechanism. So, in the chiral symmetry limit when the up
and down quark masses vanish, all the quark mixing angles
vanish. This mechanism is not dependent on the number
of quark-lepton families nor on any ``vertical'' symmetry structure,
unifying quarks and leptons inside a family as in Grand Unified
Theories. Together with a hypothesis of maximal CP violation, the
model leads to a completely predictive ansatz for all the CKM matrix
elements in terms of the quark masses. Some implications for neutrino
masses and oscillations are briefly discussed.

\thispagestyle{empty}
\newpage

\section{Introduction}

The pattern of flavour mixing and its relation to the quark-lepton
masses is one of the major outstanding problems of particle physics
(for a recent review see \cite{review}).
Many attempts have been made to interpret this pattern in terms
of various family symmetries---discrete or continuous, global or local.
Among them, the abelian $U(1)$ \cite{fn1,leurer,ibanezross}
and/or non-abelian $SU(2)$  \cite{volkas,dine,barbieri}
and $SU(3)$ \cite{jon,gerry}
chiral family symmetries seem the most promising.  They provide
some guidance to the expected hierarchy between the elements of the
quark-lepton mass matrices and to the presence
of texture zeros \cite{rrr} in them,
leading to relationships between the mass and mixing parameters.
In the framework of the supersymmetric Standard Model, such a family
symmetry should at the same time provide an almost uniform mass spectrum
for the superpartners with a high degree of flavour conservation
\cite{nilles} that makes their existence even more necessary in the
SUSY case.

Despite some progress in understanding the flavour mixing problem, one
has the uneasy feeling that, in many cases, the problem seems just to
be transferred from one place to another. The peculiar quark-lepton
mass hierarchy is replaced by a peculiar set of $U(1)$ flavour
charges or a peculiar hierarchy of Higgs field VEVs
in the non-abelian symmetry case. As a result
there are not so many distinctive and testable generic
predictions, strictly
relating the flavour mixing angles to the quark-lepton masses.

A commonly accepted framework for discussing the flavour problem
is based on the picture that, in the absence of flavour mixing, only
the particles belonging to the third generation $t$, $b$ and $\tau$
have non-zero masses. All other (physical) masses and
the mixing angles appear
as a result of the tree-level mixings of
families, related to some underlying family symmetry breaking.
They might be proportional to powers of some small
parameter $\lambda$, which are
determined by the dimensions of the family symmetry allowed operators
that generate the effective (diagonal and off-diagonal) Yukawa
couplings for the lighter families in the framework of the (ordinary or
supersymmetric) Standard Model.

Using a similar philosophy, we here suggest another possibility:
\begin{itemize}
\item Flavour mixing is correlated with the mechanism for generating
the masses of the first family and is completely absent in the
chiral symmetry limit $m_u = m_d = 0$  (and $m_{e}=0)$.
Therefore, the masses (more precisely
any of the diagonal elements of the quark and
charged lepton mass matrices)
of the second and third families are practically the same in
the gauge (unrotated) and physical bases.
\end{itemize}

This simple picture can readily be generalised to any number of
quark-lepton families, and does not depend on a ``vertical''
symmetry structure unifying quarks and leptons inside a family
as in Grand Unified Theories.
In section 2 we present the model motivated by this picture, and
the resulting predictions for the quark mixing angles in terms
of the quark masses.
Assuming a maximal form of CP violation, the CKM matrix is shown to be
completely determined and is numerically evaluated in section 3.
The implications for lepton mixing and neutrino
oscillations are briefly discussed in section 4. Finally we
present our conclusions in section 5.

\section{The Model}

The proposed flavour mixing mechanism driven solely by the
generation of the lightest family mass (hereafter called the Lightest
Flavour Mixing mechanism or LFM mechanism) could actually be realized
in two generic ways.

The first way is when the lightest family mass ($m_u$, $m_d$ or $m_e$)
appears as a result of the complex flavour mixing of all three families.
It ``runs along the main diagonal'' of the corresponding
$3 \times 3$ mass
matrix M, from the basic dominant element $M_{33}$ to the element
$M_{22}$ (via a rotation in the 2-3 sub-block of M) and then to the
primordially texture zero element $M_{11}$
(via a rotation in the 1-2 sub-block).
The direct flavour mixing of the first and third families of quarks
and leptons is supposed to be absent or negligibly small in $M$.

The second way, on the contrary, presupposes direct flavour mixing
of just the first and third families. There is no involvement of the
second family in the mixing.
In this case, the lightest mass appears in the
primordially texture zero $M_{11}$ element ``walking round the corner" (via a
rotation in the 1-3 sub-block of the mass matrix M). Certainly, this
second version of the LFM mechanism cannot be used for both the up
and the down quark families simultaneously, since mixing with the
second family members is a basic part of the CKM phenomenology
(Cabibbo mixing, non-zero $V_{cb}$ element, CP violation). However,
this second way could work for the up quark family
provided that the down quarks follow the first way.

So, there are two scenarios for the LFM mechanism to be considered.

\subsection*{Scenario A: ``$m_u$ and $m_d$ running
along the diagonal"}

We propose that the three mass matrices for the Dirac
fermions---the up quarks ($U$ = $u$, $c$, $t$), the down
quarks ($D$ = $d$, $s$, $b$) and charged leptons
($E$ = $e$, $\mu$, $\tau$)---in the Standard Model, or
supersymmetric Standard Model, are each hermitian with
three texture zeros of the following form:

\begin{equation}
M_i = \pmatrix{  0        &  a_i      &   0  \cr
	      a_i^{\ast} &   A_i     & b_i  \cr
		0        & b_i^{\ast}& B_i \cr} \qquad
i = U, \ D,\ E
\label{LFM1}
\end{equation}
It is, of course, necessary to assume some hierarchy between the
elements, which we take to be:
$B_i \gg A_i \sim \left| b_i \right| \gg  \left| a_i \right|$.
We do not attempt to derive this hierarchical structure here.
The zeros in the $\left( M_i \right)_{11}$ elements correspond
to our, and the commonly accepted, conjecture that the lightest
family masses appear as a direct result of flavour mixings. The
zeros in $\left( M_i \right)_{13}$ mean that only minimal
``nearest neighbour'' interactions occur, giving a tridiagonal
matrix structure.

Now our main hypothesis, that the second and third family diagonal mass
matrix elements are practically the same in the gauge and physical
quark-lepton bases, means that :
\begin{equation}
\vec{B} = (m_t,\ m_b,\ m_{\tau}) + \vec{\delta}
\qquad
\vec{A} = (m_c,\ m_s,\ m_{\mu}) + \vec{\delta^{\prime}}
\label{BA}
\end{equation}
Here we use the vector notation
$\vec{A} = (A_U,\ A_D,\ A_E)$ etc. The components $\delta_i$ and
$\delta_i^{\prime}$  are supposed to be much less than the
masses of the particles in the next lightest family, meaning
the second and first families respectively:
\begin{equation}
|\vec{\delta}| \ll (m_c,\ m_s,\ m_{\mu})
\qquad
|\vec{\delta^{\prime}}| \ll (m_u,\ m_d,\ m_e)
\label{deldelp}
\end{equation}
Since the trace and determinant of the hermitian matrix $M_i$ gives
the sum and product of its eigenvalues, it follows that
\begin{equation}
\vec{\delta} \simeq -(m_u,\ m_d,\ m_e)
\label{del}
\end{equation}
while the $\delta_i^{\prime}$ are vanishingly small and can
be neglected in further considerations.

It may easily be shown that our hypothesis and related equations
(\ref{BA} - \ref{del}) are entirely equivalent to the
condition that the diagonal elements ($A_i$, $B_i$), of the mass matrices
$M_i$, are proportional to the modulus square of the off-diagonal
elements ($a_i$, $b_i$):
\begin{equation}
\frac{A_i}{B_i} = \left| \frac{a_i}{b_i} \right|^2
\qquad
i = U, \ D,\ E
\label{ABab}
\end{equation}
In this paper we leave aside the question of deriving this
proportionality condition, eq.~(\ref{ABab}), from some underlying
theory beyond the Standard Model (see discussion in section \ref{con})
and proceed to calculate expressions
for all the elements of the matrices $M_i$ and the
corresponding CKM quark mixing matrix, in terms of the physical
masses.

Using the conservation of the trace, determinant and
sum of principal minors of the hermitian matrices $M_i$
under unitary transformations, we are led to a
complete determination of the moduli of all their elements.
The results can be expressed to high accuracy as follows:
\begin{equation}
\vec{A} = (m_c,\ m_s,\ m_{\mu})
\qquad
\vec{B} = (m_t - m_u,\ m_b - m_d,\ m_{\tau} - m_e)
\end{equation}
\begin{equation}
\left| \vec{a} \right| = (\sqrt{m_u m_c},\
\sqrt{m_d m_s},\ \sqrt{m_e m_{\mu}})
\end{equation}
\begin{equation}
\left| \vec{b} \right| = (\sqrt{m_u m_t},\
\sqrt{m_d m_b},\ \sqrt{m_e m_{\tau}})
\end{equation}
As to the CKM matrix $V$, we must first choose a
parameterisation appropriate to our picture of flavour mixing.
Among many possible ones, the original Euler parameterisation
recently advocated \cite{review,fritzsch1} is most convenient:
\begin{eqnarray}
V & = & \pmatrix{ c_U  &  s_U &  0  \cr
		 -s_U  &  c_U &  0  \cr
		   0   &   0  &  1  \cr}
\pmatrix{ e^{-i\phi} &  0  &  0  \cr
	    0        &  c  &  s  \cr
	    0        & -s  &  c  \cr}
\pmatrix{c_D  &  -s_D &  0  \cr
	 s_D  &  c_U &   0  \cr
	  0   &   0  &  1  \cr} \\
 & = & \pmatrix{ s_Us_Dc + c_Uc_D e^{-i\phi} &
s_Uc_Dc - c_Us_D e^{-i\phi} & s_Us \cr
c_Us_Dc - s_Uc_D e^{-i\phi} &
c_Uc_Dc + s_Us_D e^{-i\phi} & c_Us \cr
-s_Ds & -c_Ds & c \cr}
\end{eqnarray}
Here $s_{U,\ D} \equiv \sin \theta_{U,\ D}$ and
$c_{U,\ D} \equiv \cos \theta_{U,\ D}$ parameterise
simple rotations $R_{12}^{U,\ D}$ between the
first and second families for the up and down quarks
respectively, while $s \equiv \sin \theta$ and
$c \equiv \cos \theta$ parameterise a rotation
between the second and third families. This
representation of $V$ takes into account the
observed hierarchical structure of the quark masses
and mixing angles. The CP violating phase is
connected directly to the first and second families
alone.

The quark mass matrices $M_U$ and $M_D$ are diagonalised
by unitary transformations which can be written in
the form:
\begin{equation}
V_U = R_{12}^U R_{23}^D \Phi^U
\qquad
V_D = R_{12}^D R_{23}^D \Phi^D
\label{VUVD}
\end{equation}
where $\Phi^{U,\ D}$ are phase matrices, depending on the
phases of the off-diagonal elements
$a_i = \left| a_i \right| e^{i\alpha_i}$ and
$b_i = \left| b_i \right| e^{i\beta_i}$:
\begin{equation}
\Phi^U = \pmatrix{e^{i\alpha_U} & 0 & 0 \cr
		  		  0 &1 & 0  \cr
		  		  0 & 0 & e^{-i\beta_U} \cr}
\qquad
\Phi^D = \pmatrix{e^{i\alpha_D} & 0 & 0 \cr
		  		  0 &1 & 0  \cr
		  		  0 & 0 & e^{-i\beta_D} \cr}
\end{equation}
The CKM matrix is defined by
\begin{equation}
V = V_U V_D^{\dagger} =
R_{12}^U R_{23}^U \Phi^U \left( \Phi^D \right)^{\ast}
\left( R_{23}^D \right)^{-1} \left( R_{12}^D \right)^{-1}
\label{CKM1}
\end{equation}
and, after a suitable re-phasing of the quark fields,
we can use the representation
\begin{equation}
 R_{23}^U \Phi^U \left( \Phi^D \right)^{\ast}
\left( R_{23}^D \right)^{-1}
= \pmatrix{ e^{-i\phi} &  0  &  0  \cr
	    	0        &  c  &  s  \cr
	    	0        & -s  &  c  \cr}
\end{equation}
The rotation matrices $R_{23}^{U,\ D}$ and $R_{12}^{U,\ D}$
for our mass matrices
are readily calculated and the CKM matrix expressed in
terms of quark mass ratios
\begin{equation}
s_U = \sqrt{\frac{m_u}{m_c}} \qquad s_D = \sqrt{\frac{m_d}{m_s}}
\end{equation}
\begin{equation}
s = \left| \sqrt{\frac{m_d}{m_b}} -
e^{i\gamma} \sqrt{\frac{m_u}{m_t}} \right|
\label{sgam}
\end{equation}
and two phases $\phi = \alpha_D -\alpha_U$ and
$\gamma = \beta_D - \beta_U$.

It follows that the Cabibbo mixing is given by the well-known
Fritzsch formula \cite{fritzsch2}
\begin{equation}
\left| V_{us} \right| \simeq
\left| s_U - s_D e^{-i\phi} \right|
= \left|\sqrt{\frac{m_d}{m_s}} -
e^{i\phi} \sqrt{\frac{m_u}{m_c}} \right|
\end{equation}
which fits the experimental value well, provided
that the CP violating phase $\phi$ is required to
be close to $\frac{\pi}{2}$. So, in the
following, we shall assume maximal CP violation in the form
$\phi = \frac{\pi}{2}$, as is suggested by
spontaneous CP violation in the framework of $SU(3)$
family symmetry \cite{jon, harvey}. The other phase
$\gamma$ appearing in $V_{cb}$ and $V_{ub}$
\begin{equation}
\left| V_{cb} \right| \simeq s
\qquad
\left| V_{ub} \right| = s_U s
\end{equation}
can be rather arbitrary, since the contribution
$\sqrt{\frac{m_u}{m_t}}$ to $s$ is relatively
small, even compared with the uncertainties coming from
the light quark masses themselves. This leads to our
most interesting prediction (with the mass ratios
calculated at the electroweak scale \cite{koide}):
\begin{equation}
\left|V_{cb} \right| \simeq \sqrt{\frac{m_d}{m_b}} = 0.038 \pm 0.007
\label{Vcb}
\end{equation}
in good agreement with the current data
$\left| V_{cb} \right| = 0.039 \pm 0.003$ \cite{parodi}.
For definiteness we shall assume the phase $\gamma$ in $s$,
see eq. (\ref{sgam}), to
be aligned with the CP violating phase $\phi$, again as
suggested by $SU(3)$ family symmetry \cite{jon, harvey},
and take $\gamma = \frac{\pi}{2}$.
This has the effect of
reducing the uncertainty in our prediction eq.~(\ref{Vcb})
from 0.007 to 0.004. Another  prediction for the ratio:
\begin{equation}
\left| \frac{V_{ub}}{V_{cb}} \right| = \sqrt{\frac{m_u}{m_c}}
\end{equation}
is quite general for models with ``nearest-neighbour"
mixing \cite {review}.

\subsection*{Scenario B: ``$m_u$ walking around the corner,
while $m_d$ runs along the diagonal"}

Now the mass matrices for the down quarks $M_D$ and charged
leptons $M_E$ are supposed to have the same form as in
eq.~(\ref{LFM1}), while the hermitian mass matrix for the up
quarks is taken to be:
\begin{equation}
M_U = \pmatrix{  0        &  0         &   c_U  \cr
	    		 0  	  &   A_U      & 0  \cr
		c_U^{\ast}        &   0        & B_	U \cr}
\label{LFM2}
\end{equation}
All the elements of $M_U$ can again be readily determined in terms
of the  physical masses as:
\begin {equation}
A_U = m_c
\qquad
B_U = m_t - m_u
\qquad
\left| c_U \right| = \sqrt{m_u m_t}
\end{equation}
The quark mass matrices are diagonalised again by unitary
transformations as in eq.~(\ref{VUVD}), provided that the matrix
$V_U$ is changed to
\begin{equation}
V_U = R_{13}^U\Phi^U
\end{equation}
where the 1-3 plane rotation of the $u$ and $t$ quarks and the
phase matrix $\Phi^U$ (depending on the phase of the element
$c_U = \left| c_U \right| e^{i \alpha_U}$) are parameterised in the
following way:
\begin{equation}
R_{13}^U = \pmatrix{  c_{13}        &  0      &   s_{13}  \cr
	     0		 &   1        & 0  \cr
		-s_{13}  & 0		  & c_{13} \cr} \qquad
\qquad
\Phi^U = \pmatrix{   e^{i \alpha_U}        &  0      &   0  \cr
	      			0		 &   1      & 0	  \cr
					0        &   0		& 1   \cr}
\end{equation}
Here $s_{13} \equiv \sin \theta_{13}$ and
$c_{13} \equiv \cos \theta_{13}$.

The structure of the CKM matrix now differs from that of
eq.~(\ref{CKM1}) as it contains the direct 1-3 plane rotation for the
up quarks:
\begin{equation}
V = V_U V_D^{\dagger} =
R_{13}^U \Phi^U \left( \Phi^D \right)^{\ast}
\left( R_{23}^D \right)^{-1} \left( R_{12}^D \right)^{-1}
\label{CKM2}
\end{equation}
although the phases and rotations associated with the down
quarks are left the same as before. This natural parameterisation
is now quite close to the standard one \cite{parodi}. The
proper mixing angles and CP violating phase (after a
suitable re-phasing of the c quark, $c \rightarrow c e^{-i\beta_D}$)
are given by the simple and compact formulae:
\begin{equation}
\left| V_{us} \right| \simeq s_{12} = \sqrt{\frac{m_d}{m_s}}
\qquad
\left| V_{cb} \right| \simeq s_{23} = \sqrt{\frac{m_d}{m_b}}
\qquad
\left| V_{ub} \right| \simeq s_{13} = \sqrt{\frac{m_u}{m_t}}
\end{equation}
and
\begin{equation}
\delta = \alpha_D + \beta_D - \alpha_U
\end{equation}
While the values of $\left| V_{us} \right|$ and
$\left| V_{cb} \right|$ are practically the same
as in scenario A and in good agreement with experiment, a new
prediction for $\left| V_{ub} \right|$ (not depending on the
value of the CP violating phase) should allow experiment to
differentiate between the two scenarios in the near
future.

\section{The CKM Matrix}

Our numerical results for both versions of our model, with a
maximal CP violating phase (see discussion in
section \ref{con}), are summarized in the following CKM
matrix:
\begin{equation}
V_{CKM} = \pmatrix{  0.975(1)  &  0.222(4)   & 0.0023(5) \ A  \cr
			       &	     & 0.0036(6) \ B  \cr
	    	     0.222(4)  &  0.975(1)   & 0.038(4)  \cr
		     0.009(2)  &  0.038(4)   & 0.999(1) \cr}
\end{equation}
The uncertainties in brackets are largely given by the uncertainties
in the quark masses. There is clearly a real and testable
difference between scenarios A and B given by the value of
the $V_{ub}$ element. Agreement with the experimental values
of the already known CKM matrix elements \cite{parodi} looks
quite impressive. The distinctive predictions for the presently
relatively poorly known $V_{ub}$ and $V_{td}$ elements should
be tested in the near future, when the B-factories start
giving results \cite{parodi}.

\section{The Lepton Sector}

The lepton mixing matrix is defined analogously to the
CKM matrix:
\begin{equation}
U = U_{\nu} U{_E}^{\dagger}
\end{equation}
Our model predicts the contributions of the charged lepton
mixings $U_E$ to the neutrino mixing angles with high accuracy:
\begin{equation}
\sin \theta_{e\mu} = \sqrt{\frac{m_e}{m_{\mu}}}
\qquad
\sin \theta_{\mu \tau} = \sqrt{\frac{m_e}{m_{\tau}}}
\qquad
\sin \theta_{e\tau} \simeq 0
\end{equation}
These small contributions to the mixing angles
from the charged lepton sector will
not markedly effect atmospheric neutrino oscillations \cite{superkam},
which appear to require essentially maximal mixing
$\sin^2 2\theta_{\mu \tau} \simeq 1$.
However $\sin \theta_{e\mu}$ could be relevant to the
small angle MSW solution to the solar neutrino problem.
Nonetheless our picture of flavour mixing disfavours some recently
suggested models (see, e.g. \cite{pati}) using a considerable
charged lepton mixing contribution to accommodate the atmospheric and
solar neutrino data.

It follows that the large neutrino mixing responsible for
atmospheric neutrino oscillations should come from the
$U_{\nu}$ matrix associated with the neutrino mass matrix.
This requires a different mass matrix texture for the neutrinos
compared to the charged fermions \cite{hall}. So, if some universal
family symmetry breaking pattern \cite{harvey}
underlying the LFM mechanism leads to a $U_{\nu}$ matrix with
small mixing angles similar to the quarks, it will be
necessary to introduce another mechanism to generate neutrino
masses. An attractive method for generating a large neutrino
mixing within the supersymmetric Standard Model is via R-parity
violating interactions, which can give radiatively induced
neutrino masses and mixing angles in just the area required
by the current observational data \cite{chun,lep}.


\section{Conclusions \label{con}}

The present observational status of quark flavour mixing, as
described by the CKM matrix elements \cite{parodi}, shows that
the third family $t$ and $b$ quarks are largely
decoupled from the lighter families. At first sight, it looks
quite surprising that not only the 1-3 ``far neighbour"
mixing (giving the $V_{ub}$ element in $V_{CKM}$) but also the
2-3 ``nearest neighbour" mixing ($V_{cb}$) happen to be small
compared with the ``ordinary" 1-2 Cabibbo mixing ($V_{us}$)
which is determined, according to common belief, by the lightest
$u$ and $d$ quarks. This led us to the idea that all
the other mixings, and primarily the 2-3 mixing, could also be
controlled by the masses $m_u$ and $m_d$ and that the
above-mentioned decoupling of the third family $t$ and $b$
quarks is determined by the square roots of the corresponding mass ratios
$\sqrt{\frac{m_u}{m_t}}$ and
$\sqrt{\frac{m_d}{m_b}}$ respectively. So, in the chiral
symmetry limit $m_u = m_d = 0$, not only CP violation
vanishes, as argued in \cite{review}, but all the flavour
mixings disappear as well.

In such a way the Lightest Flavour Mixing (LFM) mechanism,
extended also to the charged lepton sector, was formulated in
section 2 with two possible scenarios A and B. We found that
the LFM mechanism reproduces well the values of the already
well measured CKM matrix elements and gives distinctive
predictions for the yet poorly known ones in both cases A and B.
One could say that, for the first time, there are compact
working formulae (especially compact in scenario B) for all
the CKM angles in terms of quark mass ratios.
The only unknown parameter is the CP violating
phase $\delta$. Taking it to be maximal
($\delta = \frac{\pi}{2}$, see below) we obtain a full
determination of the CKM matrix that is numerically summarized in
section 3. The LFM mechanism extended to the lepton sector
leads, as discussed in section 4, to very small
mixings from the charged leptons. If the tree level contributions to
the mixings from the neutrino sector are also small, as
suggested by the $SU(3)$ family symmetry breaking picture
linked to LFM in \cite{harvey}, it is necessary to introduce
another neutrino mass generation mechanism to accommodate
the recent Super-Kamiokande date \cite{superkam}. Within the
supersymmetric Standard Model, this could well be through
radiatively induced neutrino masses and mixings.

{}From the theoretical point of view the LFM mechanism is
based on the generic proportionality condition eq.~(\ref{ABab})
between diagonal and off-diagonal elements of the mass matrices.
For the N family case, this condition could be expressed as:
\begin{equation}
M_{22} : M_{33} : ... :M_{NN} = \left| M_{12} \right|^2 :
\left| M_{23} \right|^2 : ... : \left| M_{N-1\ N} \right|^2
\label{ABabN}
\end{equation}
showing clearly that the heavier families (4th, 5th, ...), had
they existed, would be more and more decoupled from the lighter
ones and from each other.

One might think that the condition eq.~(\ref{ABabN}) suggests
some underlying flavour symmetry, probably non-abelian $SU(N)$,
treating the N families in a special way. Indeed, for $N = 3$
families, we have found \cite{harvey} that the $SU(3)$ chiral
family (or horizontal) symmetry \cite{jon}, properly interpreted
in terms of the symmetry breaking vacuum configurations, 
can lead to
the basic condition eq.~(\ref{ABab}) for the mass matrices of the
up and down quarks and charged leptons, depending on the 
horizontal scalar fields taken.

At the same time the symmetry-breaking horizontal scalar fields,
triplets and sextets of  $SU(3)$, develop in general complex VEVs
and in cases linked to the LFM mechanism transmit a maximal CP
violating phase $\delta = \frac{\pi}{2}$ to the effective Yukawa
couplings \cite{harvey}.
Apart from the direct predictability of $\delta$ (which
was used in the numerical analysis of the CKM matrix given in
section 3), the possibility that CP symmetry is broken
spontaneously like other fundamental symmetries of the
Standard Model seems very attractive---both aesthetically and
because
it gives some clue to the flavour part of strong CP violation.
On the other hand,
spontaneous CP violation means that the scale of the $SU(3)$ family
symmetry must be rather high (not much less than $M_{GUT}$) in
order to avoid the standard domain wall problem by the
well-known inflation mechanism \cite{nilles}.

So, an $SU(3)$ family symmetry seems to be a good candidate for the
basic theory underlying our proposed LFM mechanism, although we
do not exclude the possibility of other interpretations as well.
Certainly, even without a theoretical derivation of
eq.~(\ref{ABab}), the LFM mechanism can be considered as a successful
predictive ansatz in its own right. Its further testing could
shed light on the underlying flavour dynamics and the way towards
the final theory of flavour.

\section*{Acknowledgements}

We should like to thank D. Sutherland for interesting discussions.
JLC is grateful to PPARC for financial support during his
visit to Glasgow University, when part of this work was done.

\end{document}